\pdfoutput=1

\documentclass[conference, compsoc, twoside,a4paper]{IEEEtran}
\usepackage{fancyhdr}
\makeatletter
\let\old@ps@headings\ps@headings
\let\old@ps@IEEEtitlepagestyle\ps@IEEEtitlepagestyle
\def\confheader#1{%
  \def\ps@IEEEtitlepagestyle{%
    \old@ps@IEEEtitlepagestyle%
    \def\@oddhead{\strut\hfill#1\hfill\strut}%
    \def\@evenhead{\strut\hfill#1\hfill\strut}%
  }%
  \ps@headings%
}
\makeatother

\ifCLASSOPTIONcompsoc
  \usepackage[nocompress]{cite}
\else
  \usepackage{cite}
\fi
\ifCLASSINFOpdf
  \usepackage[pdftex]{graphicx}
  \DeclareGraphicsExtensions{.pdf,.jpeg,.png,.eps}
\else
\fi

\usepackage{enumitem}
\usepackage{algorithm}
\usepackage{algorithmic}
\usepackage{color}
\usepackage{cite}
\usepackage{amsmath}
\usepackage{epstopdf}
\usepackage{booktabs}
\usepackage{multirow}

\def\eg{\textit{e.g.}}
\def\ie{\textit{i.e.}}

\def\etal{\textit{et al.}}

\newcommand*\rot{\rotatebox{90}}
\begin{document}

%

\title{Discovering Gender Differences in Facial Emotion Recognition \\ via Implicit Behavioral Cues}


\author{\IEEEauthorblockN{Maneesh  Bilalpur\IEEEauthorrefmark{1}, Seyed  Mostafa Kia\IEEEauthorrefmark{2}\IEEEauthorrefmark{3}, Tat-Seng  Chua\IEEEauthorrefmark{4}, Ramanathan  Subramanian\IEEEauthorrefmark{5}\IEEEauthorrefmark{6}}

\IEEEauthorblockA{\IEEEauthorrefmark{1}International Institute of Information Technology, Hyderabad, Telangana, India}
\IEEEauthorblockA{\IEEEauthorrefmark{2}Donders Institute for Brain, Cognition and Behaviour, Radboud University, Nijmegen, The Netherlands}
\IEEEauthorblockA{\IEEEauthorrefmark{3}Department of Cognitive Neuroscience, Radboud University Medical Centre, Nijmegen, The Netherlands}
\IEEEauthorblockA{\IEEEauthorrefmark{4}School of Computing, National University of Singapore, Singapore}
\IEEEauthorblockA{\IEEEauthorrefmark{5}Advanced Digital Sciences Center, University of Illinois at Urbana-Champaign, Singapore}
\IEEEauthorblockA{\IEEEauthorrefmark{6}School of Computing Science, University of Glasgow Singapore, Singapore}}

\maketitle

\begin{abstract}
We examine the utility of implicit behavioral cues in the form of EEG brain signals and eye movements for gender recognition (GR) and emotion recognition (ER). Specifically, the examined cues are acquired via low-cost, off-the-shelf sensors. We asked 28 viewers (14 female) to recognize emotions from unoccluded (\textit{no mask}) as well as partially occluded (\textit{eye} and \textit{mouth masked}) emotive faces. Obtained experimental results reveal that (a) reliable GR and ER is achievable with EEG and eye features, (b) differential cognitive processing especially for negative emotions is observed for males and females and (c) some of these cognitive differences manifest under partial face occlusion, as typified by the \textit{eye} and \textit{mouth mask} conditions.  
\end{abstract}


%
\IEEEpeerreviewmaketitle

\section{Introduction}\label{Intro}
The need to account for gender differences in interaction design and computing has spurred the evolution of the Gender HCI field\cite{Susan2006}. Being able to recognize user demographics such as gender can benefit interactive and gaming systems in terms of a) visual and interface design\cite{Czerwinski2002}, (b) recommending
the right games and products (via ads), and (c) providing the right motivation and feedback for enhancing user satisfaction and experience\cite{Schwark2013}. Most gender recognition (GR) systems are face or voice-based, which pose privacy concerns as face or voice samples are \textit{biometrics} which enable discovery of a person's identity.


In contrast, this work examines GR from \textit{implicit user behavioral signals} such as eye movements and EEG responses. Implicit behavioral signals are invisible to the outside world and cannot be recorded without the user's cooperation making them privacy compliant. Specifically, we attempt GR and emotion recognition (ER) captured via low-cost, off-the-shelf devices such as the \textit{Emotiv} EEG and the \textit{Eyetribe} eye tracker. While being easy-to-use, these devices have poor signal quality which makes analysis challenging.   

Specifically, we examine gender differences by posing a facial emotion recognition task. To this end, we presented 28 viewers (14 female) with facial emotions from the Radboud face database, and asked them to recognize the presented emotion. In addition to presenting viewers with unoccluded emotive faces (\textit{no mask}), we also presented them images where facial features were partially occluded, \ie, either the eyes were completely covered (\textit{eye mask}), or the mouth and lower nasal region were covered (\textit{mouth mask}) (see Fig.\ref{proto}). 

The study yielded interesting results. Firstly, examination of explicit viewer responses showed that women were more accurate at recognizing especially negative emotions. Then, investigation of EEG event related potentials (ERPs) revealed differences in ERP components for positive and negative emotions in females. Analysis of eye movements showed that females were prone to fixating on the eyes for discovering emotional cues. Finally, extensive ER and GR experiments showed that (a) reliable GR (peak AUC of 0.71) and ER (peak AUC of 0.64 for recognizing positive vs negative emotions) is achievable with the considered modalities and sensors; (b) differential cognitive processing for males and females is observable while processing negative facial emotions, and (c) some of these differences manifest under occlusion as typified by the \textit{eye mask} and \textit{mouth mask} conditions. The next section positions our work with respect to the literature in order to highlight the motivation for this study and its novelty.

\section{Related work}
Many works have focused on ER with implicit behavioral signals~\cite{zheng2014multimodal,liu2016multimodal,abadi2015decaf,Koelstra2012,subramanian2016ascertain}, but very few works have examined on GR with such signals~\cite{wu2015human}. Also, some works have attempted to capture emotion and gender-specific differences by examining eye movement and EEG responses to emotional faces~\cite{schurgin,zotto2015processing,katti2010making,Subramanian2014}. However, none of these studies have attempted to isolate behavioral differences as well as exploit them in a computational setting. A closely related work~\cite{maneesh2017icmi} exploits gender sensitivity to mild and intense facial emotions for GR. We explicitly show gender-specific EEG and eye movement patterns characteristic of emotional face processing under occluded and unoccluded conditions, and demonstrate recognition results consistent with these findings. Spatio-temporal analysis of EEG data also reveals cognitive differences while processing facial emotions in the \textit{no mask} and \textit{mask} conditions.  

In an exhaustive survey of GR methods~\cite{wu2015human}, the authors attempt to address the GR problem via metrics like universality, distinctiveness, permanence and collectability. While bio-signals like EEG can be accurate and reliable, acquiring them is nevertheless invasive requiring express cooperation from users. The use of low-cost, wearable devices makes our experimental design minimally invasive, our study ecologically valid and our framework suitable for user profiling at scale. 

Some works have expressly attempted to recognize emotions using machine learning approaches from eye/EEG data or a combination of both. Zheng \etal~\cite{zheng2014eeg} integrate deep belief networks with a hidden Markov model for distinguishing positive and negative emotions. Unsupervised ER from raw EEG is proposed in~\cite{jirayucharoensak2014eeg}. Zheng \textit{et al}~\cite{zheng2014multimodal} propose ER using differential entropy features based on EEG and pupil diameter. Liu \textit{et al}\cite{liu2016multimodal} perform ER with differential autoencoders, and attempt cross-modal ER based on shared EEG and eye-based representations. 
Tavakoli \textit{et al}~\cite{Tavakoli15} perform valence (+ve vs -ve emotion) recognition using eye movements with an emphasis on evaluating eye-based features and their fusion, and achieve 52.5\% accuracy with discriminative feature selection and a linear SVM.



Upon examining the literature, we summarize the research contributions of this work as follows: (1) It is one of the very few works to computationally perform GR and ER based on EEG and eye-based features; we also supplement the recognition results with explicit user behavioral data and isolation of EEG and eye-based patterns; (2) We use low-cost, off-the-shelf EEG and eye movement sensors to this end as against lab-grade equipment used by most works, which affirm the ecological validity of our findings and the feasibility of employing the proposed framework for large-scale user profiling; (3) Use of the \textit{mask} vs \textit{no mask} paradigms allows for a fine-grained examination of gender differences. Some gender-specific behaviors (\eg, proclivity to fixate on the eyes for women or gender information encoded in the EEG data at specific electrodes) manifest under the \textit{eye} and \textit{mouth mask} conditions. The following section describes our experimental set up and protocol. 

\section{Materials and Methods}\label{M&Ms}
\begin{figure}[t]
   \includegraphics[width=0.95\linewidth]{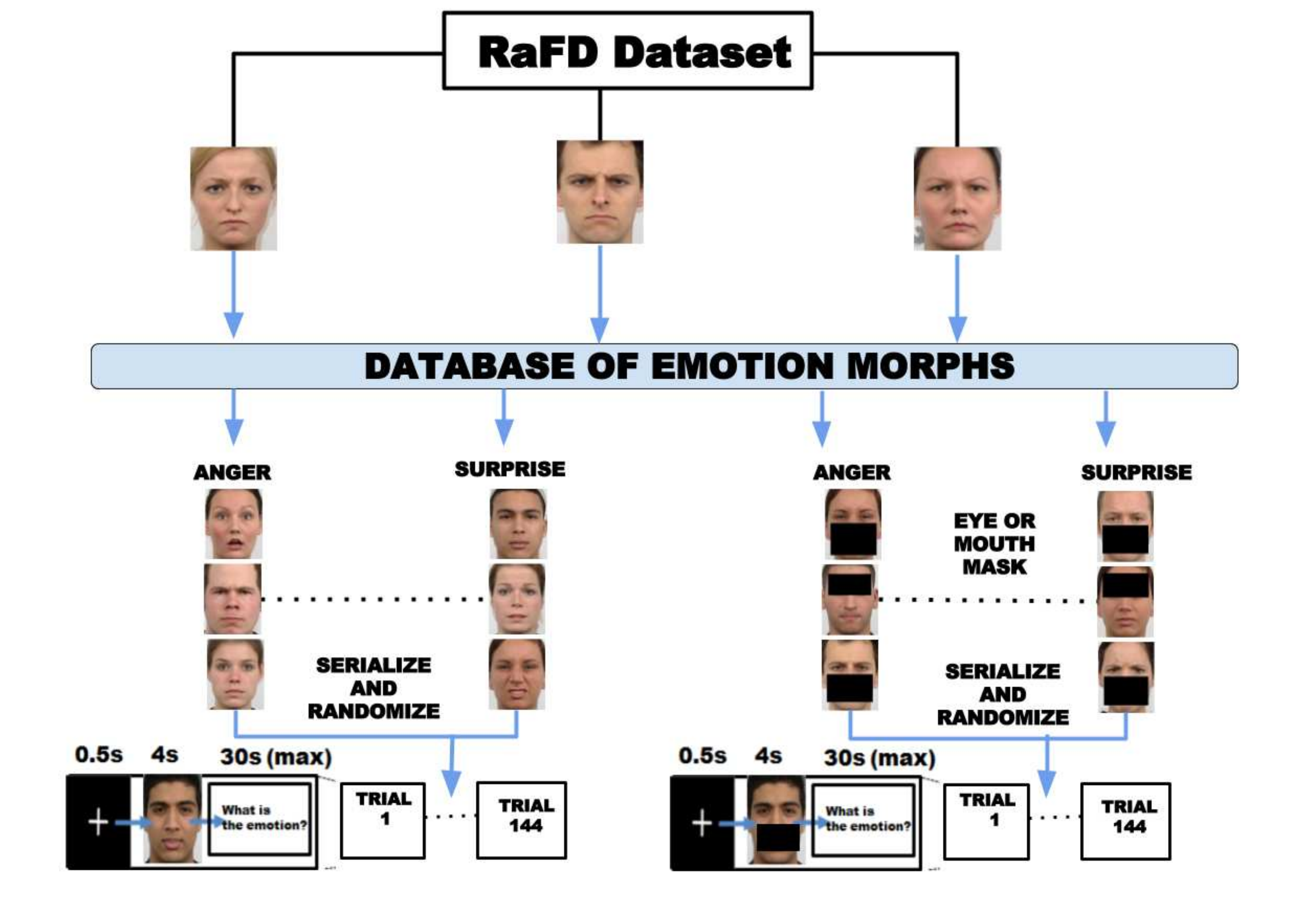} \vspace{-0.4cm}
    \caption{\label{proto} \textbf{Experimental Protocol:} Viewers were required to recognize the facial emotion from either an \textit{unmasked} face, or with the \textit{eye}/\textit{mouth} masked. Trial timelines for the two conditions are shown at the bottom.} \vspace{-0.5cm}
\end{figure}

We examined gender differences in visual emotional face processing when a) the faces were fully visible, and (b) some facial features were occluded. Specifically, we investigated gender sensitivity to emotions when the \textit{eye} and \textit{mouth} regions are occluded (see Fig.\ref{proto}), as the importance of these features towards conveying facial emotions has been highlighted by prior works~\cite{subramanian2011can,schurgin}. The three conditions are denoted as the \textit{\textbf{no mask}}, \textit{\textbf{eye mask}} and \textit{\textbf{mouth mask}} conditions throughout the paper. \\

\noindent \textbf{Participants:} 28 subjects of different nationalities (14 male, age $26.1\pm7.3$ and 14 female, age $25.5\pm6$), with normal or corrected vision, took part in our study. \\

\noindent \textbf{Stimuli:} We used emotional faces of 24 models (12 male, 12 female) from the Radboud Faces Database (RaFD)~\cite{Rafd}. RaFD includes facial emotions of 49 models rated for \textit{clarity}, \textit{genuineness} and \textit{intensity}, and the 24 models were chosen such that their Ekman facial emotions\footnote{\textbf{A}nger, \textbf{D}isgust, \textbf{F}ear, \textbf{H}appy, \textbf{S}ad and \textbf{Su}rprise.} were roughly matched based on these ratings. We then \textit{morphed} the emotive faces from \textit{neutral} (0\% intensity) to \textit{maximum} (100\% intensity) to generate intermediate morphs in steps of 5\%. Emotional morphs with 55--100\% intensity were used in this work. \textit{Eye} and \textit{mouth-masked} faces were generated upon automatically locating facial landmarks via~\cite{Baltrusaitis2016}. The \textit{eye mask} covered the eyes and nasion, while the \textit{mouth mask} covered the mouth and the nose ridge. The (un)masked faces were 361$\times$451 pixels in size, encompassing a visual angle of 9.1$^\circ$ and 11.4$^\circ$ about $x$ and $y$ at 60 cm screen distance. \\

%
\noindent \textbf{Protocol:} The protocol is outlined in Fig.\ref{proto}, and 
involved the presentation of \textit{unmasked} and \textit{masked} faces to viewers over two separate sessions, with a break in-between to avoid fatigue. We chose one face per model and emotion, resulting in 144 face images (1 morph/emotion $\times$ 6 emotions $\times$ 24 models). In the first session, these faces were shown as is in random order (no mask condition), and were re-presented with an eye or mouth mask (eye/mouth mask conditions) in the second session. We ensured a 50\% split of the eye/mouth-masked faces, and the masked faces were also shown randomly to viewers.

During each trial, the (un)masked face was displayed for 4s preceded by a fixation cross for 500 ms. The viewer then had a maximum of 30s to make {one} out of {seven} choices concerning the facial emotion (6 emotions plus neutral) via a radio button. Neutral faces were only utilized for morphing purposes and not used in the experiment. During the experiment, viewers' EEG signals were acquired via the 14-channel \textit{Emotiv epoc} device, and eye movements were recorded with the \textit{Eyetribe} tracker. The experiment was split into 4 segments to minimize acquisition errors, and took about 90 minutes. 

\subsection{User Responses}\label{UR}
We first compare male and female sensitivity to emotions based on response times (RTs) and recognition rates (RRs), and then proceed to examine their implicit eye movements and EEG responses. RT and RR denote performance measures computed from \textit{explicit} viewer responses. \\

\begin{figure}[t]
\includegraphics[width = 1\linewidth]{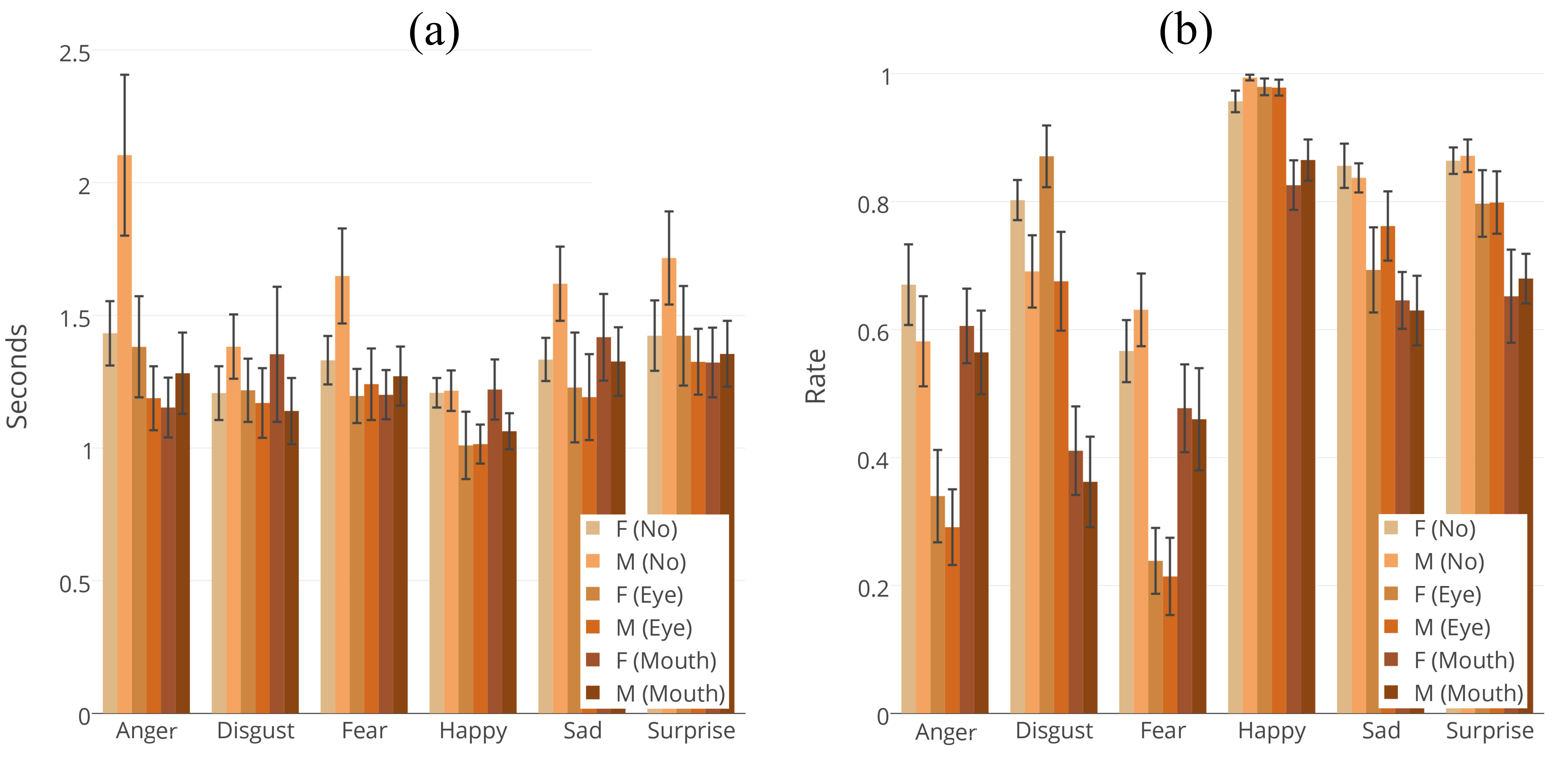} \vspace{-0.3cm}
\caption{ (a) RTs and (b) RRs for the three conditions. Error-bars denote unit standard error. View in color and under zoom.} \vspace{-0.5cm}
\label{RTnRR}
\end{figure}

\noindent \textbf{Response Time (RTs):} Overall RTs for the \textit{no mask}, \textit{eye mask} and \textit{mouth mask} conditions were respectively found to be 1.44$\pm$0.24, 1.17$\pm$0.12 and 1.25$\pm$0.09 seconds, implying that facial ER was fairly instantaneous, and that viewer responses were surprisingly faster with masked faces. A fine-grained comparison of male and female RTs in the three conditions (Fig.\ref{RTnRR}(a)) revealed that females ($\mu_{\text{RT}} = 1.40\pm 0.10$s) were generally faster than males ($\mu_{\text{RT}} = 1.60\pm 0.10$s) at recognizing \textit{unmasked} emotions. Female alacrity nevertheless decreased for masked faces, with males responding marginally faster for \textit{eye masked} faces ($\mu_{\text{RT}}\text{(male)} = 1.13\pm0.11$s vs $\mu_{\text{RT}}\text{(female)} = 1.21\pm0.13$s), and both genders responding with similar speed for \textit{mouth masked} faces ($\mu_{\text{RT}}\text{(male)} = 1.24\pm0.10$s vs $\mu_{\text{RT}}\text{(female)} = 1.25\pm0.09$s). \\

\noindent \textbf{Recognition Rates (RRs):} Overall, RRs for \textit{unmasked} emotions ($\mu_{\text{RR}}=77.6$) were expectedly higher than for \textit{eye-masked} ($\mu_{\text{RR}}=59.7$) and mouth-masked ($\mu_{\text{RR}}=63.5$) emotions. Happy faces were recognized most accurately in all three conditions. Specifically focusing on gender differences (Fig.\ref{RTnRR}(b)), females recognized facial emotions more accurately than males and this was particularly true for negative (A,D,F,S) emotions-- male vs female RRs for these emotions differed significantly in the \textit{no mask} ($\mu_{\text{RR}}\text{(male)} =54.3$ vs $\mu_{\text{RR}} \text{(female)} =61.2, p<0.05$) and \textit{eye mask} conditions ($\mu_{\text{RR}}\text{(male)} =51.8$ vs $\mu_{\text{RR}} \text{(female)} =58.1, p<0.05$), and marginally for the \textit{mouth mask} condition 
($\mu_{\text{RR}}\text{(male)} =52$ vs $\mu_{\text{RR}} \text{(female)} =55.8, p=0.08$). 

\section{Analyzing Implicit Behavioral Cues}\label{IBC}
We first describe the methodology employed to extract brain and eye-related features, before discussing emotion and gender recognition with the extracted features. 

\subsection{EEG preprocessing}
We extracted epochs for each trial (4.5s of stimulus-plus-fixation viewing time @ 128 Hz sampling rate), and the 64 leading pre-stimulus samples were used to remove DC offset. This was followed by (a) EEG band-limiting to within 0.1--45 Hz, (b) Removal of noisy epochs via visual inspection, and (c) Independent component analysis (ICA) to remove artifacts relating to eye-blinks, and eye and muscle movements. Muscle movement artifacts in EEG are mainly concentrated in the 40--100 Hz band, and are removed upon band-limiting and via inspection of ICA components. Finally, a 7168 dimensional (14$\times$4$\times$128) EEG feature vector was generated from concatenation of the 14 EEG channels over 4s of stimulus viewing.

\subsubsection{ERP analysis}\label{ERP}   
Event Related Potentials (ERPs) are (averaged) time-locked neural responses related to sensory and cognitive events. ERPs occurring in the first 100ms post stimulus presentation are stimulus-related (or exogenous), while later ERPs are cognition-related (or endogenous). We examined the first second of EEG epochs for ERPs characteristic of emotion and gender.

\begin{figure}[t]
    \centering
	  \includegraphics[width = 1\linewidth]{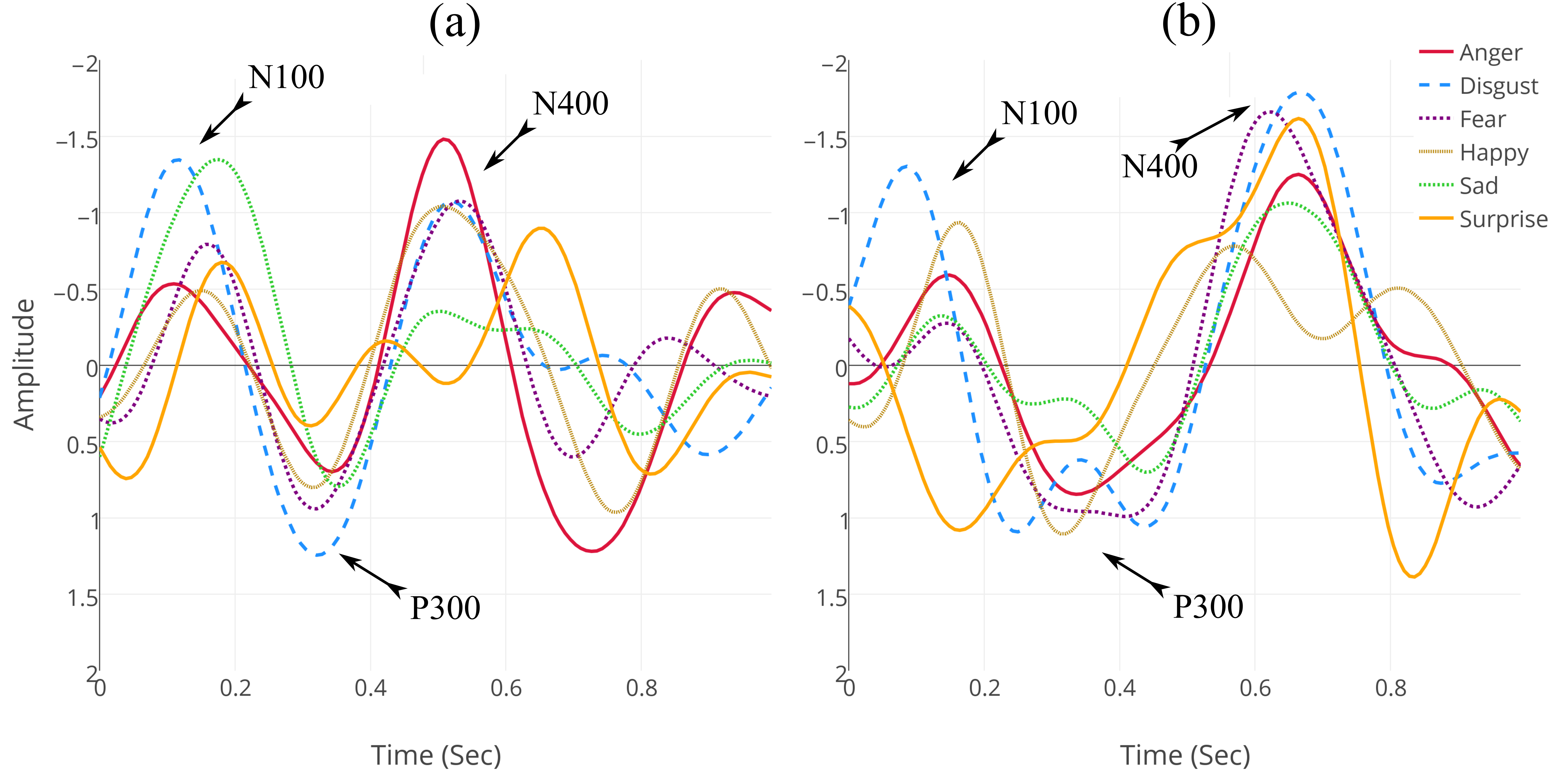} \vspace{-0.6cm}
    \caption{Female ERPs in the \textit{no mask} (a) and \textit{eye mask} (b) conditions. Best-viewed in color and under zoom.} \vspace{-0.5cm}
    \label{ERPs1}
\end{figure}

In a related work studying gender differences in emotional processing, Lithari \etal~\cite{Lithari2010} observed enhanced N100 and P300 ERP components in females for negative stimuli. 
Fig.\ref{ERPs1} shows female ERPs\footnote{ERP data is plotted upside down as per convention.} observed in the occipital O2 electrode for the \textit{no mask} and \textit{eye mask} conditions. The occipital lobe represents the visual processing center in the brain. Clearly, one can note enhanced N100 and P300 ERP components for negative emotions in the \textit{no mask} condition (Fig.\ref{ERPs1}(a)). This effect is attenuated in the \textit{eye mask} (Fig.\ref{ERPs1}(b)) and \textit{mouth mask} conditions, although one can note stronger N400 amplitudes for anger and disgust in the \textit{eye mask} case. Such a pattern was not observable from male ERPs. Overall, the observed ERPs affirm that gender differences in emotional processing can be captured with the low-cost Emotiv EEG device. 

\subsection{Gaze data preprocessing}
\begin{figure*}[t]
\centering
\includegraphics[width = 0.825\linewidth]{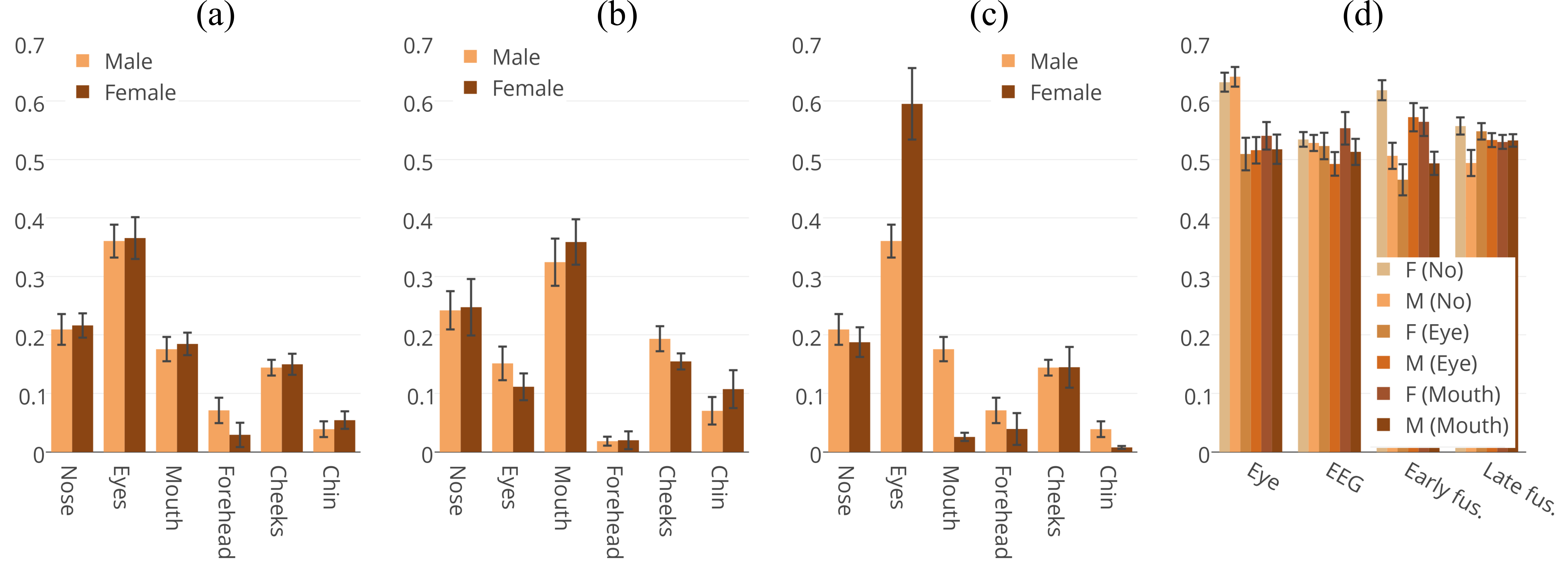} \vspace{-0.3cm}
\caption{(a--c) Fixation duration distributions for males and females in the \textit{no-mask}, \textit{eye-masked} and \textit{mouth-masked} conditions. (d) Valence recognition with various modalities. Best-viewed in color and under zoom.}\vspace{-0.5cm}
\label{Fix_dur_dist}
\end{figure*}

Raw gaze data acquired from the \textit{EyeTribe} device at 30 Hz sampling rate were used to extract \textit{fixations} with 100 ms temporal threshold via the \textit{EyeMMV} toolbox. \textit{Saccades} were assumed as the transitions between fixations. We also computed the duration fixated by male and female viewers for the six facial regions, namely, nose, eyes, mouth, forehead, cheeks and chin.
The normalized fixation duration distribution over these six regions for the three conditions is presented in Fig.\ref{Fix_dur_dist}(a--c).

Fig.\ref{Fix_dur_dist}(a--c) show that the eye, mouth and nasal face regions attract maximum visual attention during facial emotion recognition, in line with prior studies such as~\cite{schurgin,subramanian2011can}. Prior works such as \cite{wells2016identification} have also observed that females look up to the eyes for emotional cues, which is mirrored by longer female fixations around the eyes in the \textit{no mask} and \textit{mouth mask} conditions. Fig.\ref{Fix_dur_dist}(b) suggests that when eye information is unavailable, females tend to concentrate on the mouth and nasal regions for emotional cues. Upon extracting the fixations and saccades, we adopted features used for recognizing \textit{valence} (pleasant vs unpleasant emotions) in~\cite{Tavakoli15}, namely, saccade orientation, top-ten salient coordinates, saliency map and histograms of a) saccade slope, b) saccade length, c) saccade velocity, d) fixation duration, e) fixation count, and f) saliency to compute the 825-dimensional feature vector.

\vspace{-0.1cm}
\subsection{Experiments and Results}
This section presents emotion recognition (ER) and gender recognition (GR) results obtained with EEG features, eye-based features and their combination, under the three considered conditions. {Recognition experiments were performed on data compiled from trials where viewers correctly recognized the presented facial emotion.}

\subsubsection{Experimental settings} We considered (i) EEG features, (ii) attention-based features, (iii) concatenation of the two (termed \textit{early fusion} or EF), and (iv) probabilistic fusion of the EEG and eye-based classifiers (\textit{late fusion} or LF) for our analyses, in the three conditions during our recognition experiments. For LF, we employed the $W_{est}$ technique~\cite{koelstra2012fusion} briefly described as follows. From the EEG and eye-based classifier outputs, the test set posterior probability for class $j$ is computed as $\sum_{i=1}^2 \alpha_i^*t_ip_i^j$, where $i$ indexes the two modalities, $p_i^j$'s denote posterior classifier probability for class  $j$ with modality $i$ and $\{\alpha_i^*\}$ are the optimal weights maximizing test performance, as determined via a 2D grid search. If $F_i$ denotes the training performance for the $i^{th}$ modality, then the normalized training performance $t_i = \alpha_i F_i/\sum_{i=1}^2 \alpha_i F_i$ for given $\alpha_i$. 

We considered four classifiers including Naive Bayes (NB), Linear Discriminant Analysis (LDA), Support Vector Machines with RBF (RSVM) and linear (LSVM)  kernels in our recognition experiments. We adopted the \textit{area under ROC curve} (AUC) metric for performance evaluation. {AUC measures area under the ROC curve, which plots the true positive rate against false positive rate with varying threshold. A random classifier would generate an AUC score of 0.5, while a perfect classifier would return an AUC of 1.} Also, as we attempt recognition with few training data, we report ER/GR results over five repetitions of 10-fold cross validation (CV). CV is typically used to overcome the \textit{overfitting} problem on small datasets, and optimal SVM parameters were determined from the range $[10^{-4},10^{4}]$ via an inner 10-fold CV on the training set.  

\subsubsection{Valence (+ve vs -ve emotion) Recognition}
As explicit behavioral responses (Sec.\ref{UR}) and ERP-based (Sec.\ref{ERP}) analyses suggest that (a) females are especially sensitive to negative facial emotions, and (b) there are cognitive differences in the processing of positive and negative emotions for females, we attempted to classify positive (H, Su) and negative (A,D,F,S) emotions.Fig.\ref{Fix_dur_dist}(d) presents ER results obtained with the LDA classifier for male and female data, with unimodal and multimodal features for the three conditions. The best ER results are expectedly achieved for the no mask condition, and with eye-based features (AUC = 0.63 for females, and 0.64 for males), while minimal recognition is noted for the \textit{eye mask} condition reiterating the importance of eyes towards encoding emotional cues. EEG-based features only produce slightly better than chance recognition with a peak AUC score of 0.54. Multimodal recognition is slightly better than unimodal, and higher AUC scores are achieved in the masked conditions suggesting the complementarity of EEG and eye-based features when only partial facial information is available. 

While observed ER results are modest, they are still better or competitive with respect to prior eye and neural based ER approaches. Eye based features are found to achieve 52.5\% valence recognition accuracy in~\cite{Tavakoli15}, where emotions are induced in viewers by presenting a diverse types of emotional scenes to viewers as against emotional faces specifically employed in this work. Also, prior neural-based emotional studies~\cite{abadi2015decaf,Muhl14} achieve only around 60\% valence recognition with lab-grade sensors. On the contrary, implicit behavioral cues are acquired with low-cost EEG and eye tracking devices in this work.


\subsubsection{Gender Recognition}
\begin{table}[bh!]
\vspace{-0.5cm}
\centering
\caption{GR with different conditions/modalities. The best classifier is denoted within parenthesis. Highest AUC for a given condition is shown in bold.}
\label{GenResults}
\resizebox{0.35\textwidth}{!}{
\begin{tabular}{l|l|c|ccc}
\multicolumn{3}{c|}{\textbf{AUC}} & \textbf{Unmasked} & \textbf{Eye mask} & \textbf{Mouth mask} \\ \hline
\multicolumn{2}{c|}{\multirow{4}{*}{\textbf{All}}} & \textbf{EEG (NB)} & \textbf{0.71$\pm$0.01} & \textbf{0.69$\pm$0.03} & \textbf{0.65$\pm$0.04} \\
\multicolumn{2}{c|}{} & \textbf{EYE (NB)} & 0.49$\pm$0.01 & 0.47$\pm$0.06 & 0.52$\pm$0.05 \\
\multicolumn{2}{c|}{} & \textbf{EF (RSVM)} & 0.52$\pm$0.03 & 0.52$\pm$0.07 & 0.52$\pm$0.06 \\
\multicolumn{2}{c|}{} & \textbf{LF (RSVM)} & 0.55$\pm$0.02 & 0.54$\pm$0.05 & 0.61$\pm$0.07 \\ \hline
\multirow{24}{*}{\rot{\textbf{Emotion Wise}}} & \multirow{6}{*}{\rot{\textbf{EEG (NB)}}} & A & \textbf{0.71$\pm$0.06} & 0.61$\pm$0.16 & \textbf{0.67$\pm$0.07} \\
 &  & D & 0.67$\pm$0.05 & 0.59$\pm$0.06 & 0.65$\pm$0.14 \\
 &  & F & 0.64$\pm$0.06 & 0.60$\pm$0.14 & 0.56$\pm$0.12 \\
 &  & H & 0.69$\pm$0.05 & 0.58$\pm$0.07 & 0.62$\pm$0.08 \\
 &  & Sa & 0.67$\pm$0.05 & \textbf{0.65$\pm$0.08} & 0.59$\pm$0.08 \\
 &  & Su & 0.69$\pm$0.05 & 0.63$\pm$0.08 & 0.65$\pm$0.09 \\ \cline{2-6} 
 & \multirow{6}{*}{\rot{\textbf{EYE (NB)}}} & A & \textbf{0.60$\pm$0.02} & 0.47$\pm$0.20 & \textbf{0.59$\pm$0.16} \\
 &  & D & 0.57$\pm$0.01 & 0.48$\pm$0.14 & 0.52$\pm$0.27 \\
 &  & F & 0.59$\pm$0.01 & \textbf{0.68$\pm$0.25} & 0.58$\pm$0.17 \\
 &  & H & 0.56$\pm$0.02 & 0.55$\pm$0.13 & 0.54$\pm$0.12 \\
 &  & Sa & 0.54$\pm$0.01 & 0.44$\pm$0.15 & 0.45$\pm$0.13 \\
 &  & Su & 0.55$\pm$0.01 & 0.62$\pm$0.13 & 0.49$\pm$0.16 \\ \cline{2-6} 
 & \multirow{6}{*}{\rot{\textbf{EF (RSVM)}}} & A & 0.55$\pm$0.02 & 0.32$\pm$0.25 & 0.39$\pm$0.18 \\
 &  & D & 0.53$\pm$0.02 & 0.52$\pm$0.18 & 0.50$\pm$0.24 \\
 &  & F & \textbf{0.62$\pm$0.01} & \textbf{0.70$\pm$0.22} & 0.52$\pm$0.18 \\
 &  & H & 0.57$\pm$0.02 & 0.57$\pm$0.13 & \textbf{0.52$\pm$0.14} \\
 &  & Sa & 0.54$\pm$0.02 & 0.53$\pm$0.16 & 0.50$\pm$0.14 \\
 &  & Su & 0.58$\pm$0.01 & 0.43$\pm$0.13 & 0.51$\pm$0.16 \\ \cline{2-6} 
 & \multirow{6}{*}{\rot{\textbf{LF (RSVM)}}} & A & 0.54$\pm$0.06 & 0.57$\pm$0.15 & 0.59$\pm$0.09 \\
 &  & D & 0.54$\pm$0.04 & 0.59$\pm$0.10 & 0.57$\pm$0.14 \\
 &  & F & 0.52$\pm$0.03 & \textbf{0.64$\pm$0.16} & \textbf{0.69$\pm$0.12} \\
 &  & H & 0.52$\pm$0.03 & 0.56$\pm$0.09 & 0.55$\pm$0.10 \\
 &  & Sa & \textbf{0.57$\pm$0.07} & 0.51$\pm$0.04 & 0.58$\pm$0.11 \\
 &  & Su & 0.56$\pm$0.07 & 0.58$\pm$0.08 & 0.58$\pm$0.08 \\ \hline
\end{tabular}}
\end{table}
Very few works have investigated GR via EEG and eye-based responses. Table~\ref{GenResults} presents GR results with features extracted for (a) \textit{all} faces, and (b) \textit{emotion-specific} faces for the three conditions. The classifier employed for GR is also specified with parentheses. As with ER, the best GR performance (AUC = 0.71) is observed for the \textit{no mask} condition but with EEG features. With both EEG and eye-based cues, slightly higher GR performance is noted for the \textit{eye mask} condition as compared to the \textit{mouth mask} condition considering viewer responses for \textit{all} as well as \textit{emotion specific} faces. Overall, EEG features appear to encode gender differences better than eye-based features. Also, early or late fusion of the two modalities does not improve GR performance by much, even though one can note that fusion performs better than at least one of the modalities in the \textit{eye mask} and \textit{mouth mask} conditions. 

A close examination of the emotion-wise results reveals that optimal GR is mainly achieved with viewer responses observed for negative facial emotions. This again is consistent with the observations from explicit user responses (Sec.\ref{UR}), where females were found to recognize negative facial emotions better than males. Overall, the obtained results point to systematic differences between males and females in the cognitive processing of emotional faces. Focusing on the standard deviation (sd) in AUC scores for emotion-wise GR, one can note 
much higher sd values in the \textit{eye mask} and \textit{mouth mask} conditions as compared to \textit{no mask}, indicating the difficulty in recognizing facial emotions under occlusion conditions. However, relatively lower sd values are noted with LF implying that probabilistic fusion enables better exploitation of the complementarity in the eye and brain-based signals, resulting in more robust GR. 
\begin{figure*}
\centering
\includegraphics[width = 0.825\linewidth]{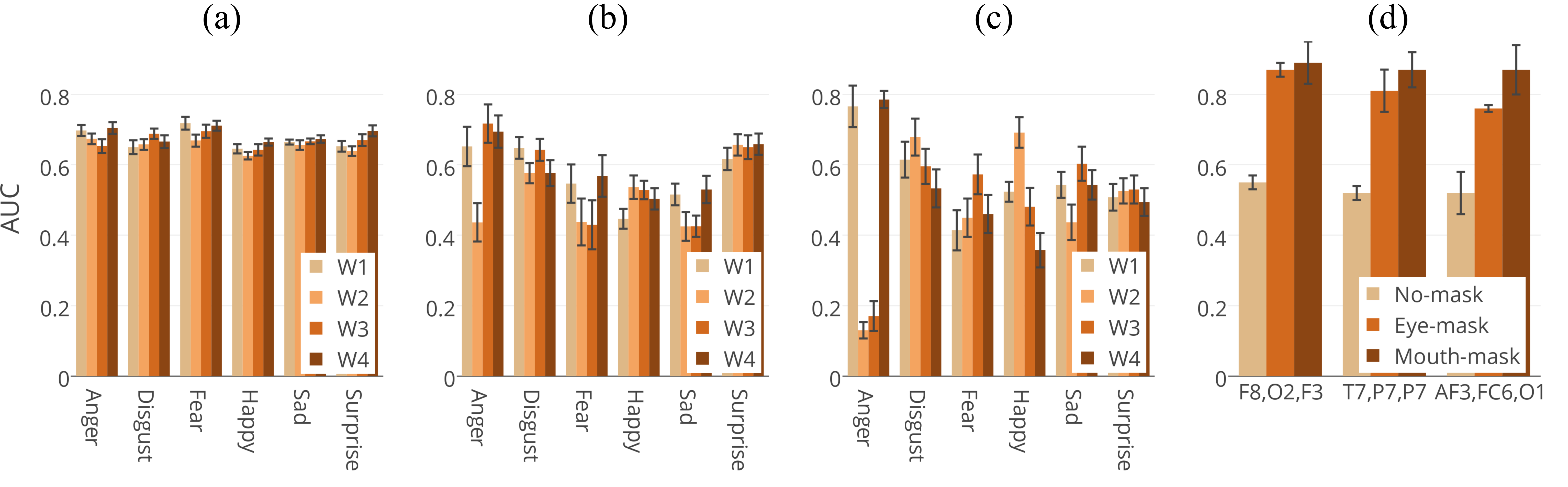} \vspace{-0.3cm}
\caption{(a--c): Temporal variance in GR performance for the \textit{no mask}, \textit{eye mask} and \textit{mouth mask} conditions. (d) Top three electrodes showing best GR performance for the three conditions.}\vspace{-0.5cm}
\label{winGRnVR}
\end{figure*}


\subsubsection{Spatio-temporal EEG analysis} As EEG represents multi-channel time-series data, we performed spatial and temporal analyses on the EEG signals to discover (i) the critical time window for GR over the 4s of stimulus presentation, and (ii) the key EEG electrodes encoding gender differences in emotional face processing.

Fig.\ref{winGRnVR}(a--c) show the temporal variance in GR performance for three considered conditions when the 4s presentation time is split into four 1s-long windows. These results are computed considering emotion-specific EEG epochs using RSVM. No significant difference in GR performance is observed over time for the \textit{no mask} scenario, and an AUC score of around 0.65 can be consistently noted over time. Barring exceptions, the GR AUC scores for the \textit{eye mask }and \textit{mouth mask} conditions are lower than for the \textit{no mask} condition. Also, maximum GR is predominantly noted for later temporal windows for mask conditions, implying that the discriminative cognitive processing between males and females happens later in time when only partial emotional information is made available to viewers. For the \textit{eye mask} scenario, maximum GR AUC scores are achieved for anger (AUC = 0.71 in W3), disgust (AUC = 0.65 in W1) and surprise (AUC = 0.65 in W2, W3 and W4). For the \textit{mouth mask} condition, highest AUCs are achieved for anger (AUC = 0.76 in W1 and 0.78 in W4), disgust (AUC = 0.68 in W2) and happy (AUC=0.69 in W2). However, low GR performance is generally noted over all time windows for the fear and surprise emotions in this condition, attributable to the general difficulty in recognizing these emotions when mouth-related information is unavailable. 

Results of spatial analysis to identify the (top three) electrodes that best encode gender differences in emotion processing\footnote{EEG data over all 4s is used for this analysis.} are shown in Fig.\ref{winGRnVR}(d). In the \textit{no mask} condition, very similar AUC scores are observed for the frontal F8 (0.55), AF3 (0.52) and T7 (0.52) electrodes. Much higher numbers are noted for the \textit{eye mask} and \textit{mouth mask} conditions, suggesting that region-specific differences in cognitive processing for males and females manifest under occlusion conditions. For the \textit{eye mask} condition, highest AUCs are noted for O2 (0.87), P7 (0.81) and FC6 (0.76), while for \textit{mouth mask} F3 (AUC = 0.89), P7 (AUC = 0.87) and O1 (AUC = 0.87). An interesting observation is that a majority of the electrodes are located on the left hemisphere of the brain (see Emotiv electrode configuration for details).
%

\vspace{-0.1cm}
\section{Discussion and Conclusion}
This work expressly examined the utility of implicit behavioral cues, in the form of EEG brain signals and eye movements to encode gender-specific information during visual processing of emotional faces. In addition to the use of unoccluded (\textit{no mask}) emotive faces, we also employed occluded (\textit{eye} and \textit{mouth masked}) faces to this end. 

The study yielded several interesting results. Firstly, examination of explicit viewer responses revealed the higher sensitivity of females to negative emotions. We then proceeded to examine their implicit behavioral signals, which revealed (a) stronger N100, P300 and N400 female ERP components for negative emotions, and (b) the proclivity of females to fixate on the eyes for discovering emotional cues.   

Emotion and gender recognition experiments with EEG and eye-based features, as well as their combination revealed that optimal recognition of emotional valence is achieved with eye-based features, while best GR performance is achieved with EEG. Also, best GR with emotion-specific features is noted for negative emotions consistent with the analysis involving explicit viewer responses. EEG features generally outperform eye-based features for the purpose of GR. Also, the fusion of EEG and eye-based features is not significantly beneficial even though more robust classification results are achieved via late fusion of unimodal results, suggesting that the EEG (neural) and eye (physiological) signals encode complementary gender-specific information. Finally, spatio-temporal analysis of EEG data revealed that superior GR is achieved with EEG from later time windows in the mask conditions, and that frontal and occipital electrodes best encode gender differences in cognitive processing of emotional faces. 

The obtained results are competitive in terms of ER and GR performance, especially considering that implicit user behavioral cues are acquired with low cost sensors, and also consistent with prior observations. Lithari \etal~\cite{Lithari2010} noted amplitude and phase-related differences for positive and negative emotions in males and females, and some of these findings are mirrored in our ERP analyses. Several eye tracking works (\eg, see~\cite{halljudith}) have noted enhanced female sensitivity for negative emotions, while other neural studies~\cite{Mclure04} have noted that gender differences in emotion processing are best encoded by the frontal and occipital brain lobes. As reliable GR (max AUC = 0.71) is achievable with the considered features, future work will explore use of these modalities for user profiling in gaming and interactive scenarios to facilitate applications like targeted advertising. {Also, recent developments in the field of deep learning can be exploited to discover latent feature spaces that are robust to noise and best capture gender-based cognitive impressions.}

\vspace{-0.1cm}
\section*{Acknowledgement}
\vspace{-0.3cm}
\noindent This study is supported by the Human Centered Cyber-physical Systems research grant from Singapore's Agency for Science, Technology and Research (A*STAR).
\tiny
\bibliographystyle{IEEEtran}
\bibliography{sample}

\begin{thebibliography}{10}
\providecommand{\url}[1]{#1}
\csname url@samestyle\endcsname
\providecommand{\newblock}{\relax}
\providecommand{\bibinfo}[2]{#2}
\providecommand{\BIBentrySTDinterwordspacing}{\spaceskip=0pt\relax}
\providecommand{\BIBentryALTinterwordstretchfactor}{4}
\providecommand{\BIBentryALTinterwordspacing}{\spaceskip=\fontdimen2\font plus
\BIBentryALTinterwordstretchfactor\fontdimen3\font minus
  \fontdimen4\font\relax}
\providecommand{\BIBforeignlanguage}[2]{{%
\expandafter\ifx\csname l@#1\endcsname\relax
\typeout{** WARNING: IEEEtran.bst: No hyphenation pattern has been}%
\typeout{** loaded for the language `#1'. Using the pattern for}%
\typeout{** the default language instead.}%
\else
\language=\csname l@#1\endcsname
\fi
#2}}
\providecommand{\BIBdecl}{\relax}
\BIBdecl

\bibitem{Susan2006}
S.~Wiedenbeck, V.~Grigoreanu, L.~Beckwith, and M.~Burnett, ``Gender {HCI}: What
  about the software?'' \emph{Computer}, vol.~39, pp. 97--101, 2006.

\bibitem{Czerwinski2002}
M.~Czerwinski, D.~S. Tan, and G.~G. Robertson, ``Women take a wider view,'' in
  \emph{Conference on Human Factors in Computing Systems (CHI)}, 2002, pp.
  195--202.

\bibitem{Schwark2013}
J.~D. Schwark, I.~Dolgov, D.~Hor, and W.~Graves, ``Gender and personality trait
  measures impact degree of affect change in a hedonic computing paradigm,''
  \emph{International Journal of Human Computer Interaction}, vol.~29, no.~5,
  pp. 327--337, 2013.

\bibitem{zheng2014multimodal}
W.-L. Zheng, B.-N. Dong, and B.-L. Lu, ``Multimodal emotion recognition using
  {EEG} and eye tracking data,'' in \emph{Engineering in Medicine and Biology
  Society (EMBC)}, 2014, pp. 5040--5043.

\bibitem{liu2016multimodal}
W.~Liu, W.-L. Zheng, and B.-L. Lu, ``Multimodal emotion recognition using
  multimodal deep learning,'' \emph{arXiv preprint arXiv:1602.08225}, 2016.

\bibitem{abadi2015decaf}
M.~Abadi, R.~Subramanian, S.~Kia, P.~Avesani, I.~Patras, and N.~Sebe, ``{DECAF:
  MEG}-based multimodal database for decoding affective physiological
  responses,'' \emph{IEEE Transactions on Affective Computing}, vol.~6, no.~3,
  pp. 209--222, 2015.

\bibitem{Koelstra2012}
S.~Koelstra, C.~M\"{u}hl, M.~Soleymani, J.-S. Lee, A.~Yazdan, T.~Ebrahimi,
  T.~Pun, A.~Nijholt, and I.~Patras, ``{DEAP: A Database for Emotion Analysis
  Using Physiological Signals},'' \emph{IEEE Transactions on Affective
  Computing}, vol.~3, no.~1, pp. 18--31, 2012.

\bibitem{subramanian2016ascertain}
R.~Subramanian, J.~Wache, M.~Abadi, R.~Vieriu, S.~Winkler, and N.~Sebe,
  ``{ASCERTAIN: E}motion and personality recognition using commercial
  sensors,'' \emph{IEEE Transactions on Affective Computing}, 2016.

\bibitem{wu2015human}
Y.~Wu, Y.~Zhuang, X.~Long, F.~Lin, and W.~Xu, ``Human gender classification: A
  review,'' \emph{arXiv preprint arXiv:1507.05122}, 2015.

\bibitem{schurgin}
M.~W. Schurgin, J.~Nelson, S.~Iida, H.~Ohira, J.~Y. Chiao, and S.~L.
  Franconeri, ``Eye movements during emotion recognition in faces,''
  \emph{Journal of Vision}, vol.~14, no.~13, pp. 1--16, 2014.

\bibitem{zotto2015processing}
M.~D. Zotto and A.~J. Pegna, ``Processing of masked and unmasked emotional
  faces under different attentional conditions: an electrophysiological
  investigation,'' \emph{Frontiers in psychology}, vol.~6, p. 1691, 2015.

\bibitem{katti2010making}
H.~Katti, R.~Subramanian, M.~Kankanhalli, N.~Sebe, T.-S. Chua, and K.~R.
  Ramakrishnan, ``Making computers look the way we look: exploiting visual
  attention for image understanding,'' in \emph{{ACM I}nt'l conference on
  Multimedia}, 2010, pp. 667--670.

\bibitem{Subramanian2014}
R.~Subramanian, D.~Shankar, N.~Sebe, and D.~Melcher, ``{Emotion modulates eye
  movement patterns and subsequent memory for the gist and details of movie
  scenes.}'' \emph{Journal of vision}, vol.~14, no.~3, pp. 1--18, 2014.

\bibitem{maneesh2017icmi}
M.~Bilalpur, S.~M. Kia, M.~Chawla, T.-S. Chua, and R.~Subramanian, ``Gender and
  emotion recognition with implicit user signals,'' in \emph{International
  Conference on Multimodal Interaction}, 2017.

\bibitem{zheng2014eeg}
W.-L. Zheng, J.-Y. Zhu, Y.~Peng, and B.-L. Lu, ``{EEG}-based emotion
  classification using deep belief networks,'' in \emph{International
  Conference on Multimedia and Expo (ICME)}, 2014, pp. 1--6.

\bibitem{jirayucharoensak2014eeg}
S.~Jirayucharoensak, S.~Pan-Ngum, and P.~Israsena, ``Eeg-based emotion
  recognition using deep learning network with principal component based
  covariate shift adaptation,'' \emph{The Scientific World Journal}, vol. 2014,
  2014.

\bibitem{Tavakoli15}
H.~R.-Tavakoli, A.~Atyabi, A.~Rantanen, S.~J. Laukka, S.~Nefti-Meziani, and
  J.~Heikkilä, ``Predicting the valence of a scene from observers’ eye
  movements,'' \emph{PLoS ONE}, vol.~10, no.~9, pp. 1--19, 2015.

\bibitem{subramanian2011can}
R.~Subramanian, V.~Yanulevskaya, and N.~Sebe, ``Can computers learn from humans
  to see better?: {I}nferring scene semantics from viewers' eye movements,'' in
  \emph{{ACM International Conference on Multimedia}}, 2011, pp. 33--42.

\bibitem{Rafd}
O.~Langner, R.~Dotsch, G.~Bijlstra, D.~H.~J. Wigboldus, S.~T. Hawk, and A.~van
  Knippenberg, ``Presentation and validation of the radboud faces database,''
  \emph{Cognition and Emotion}, vol.~24, no.~8, pp. 1377--1388, 2010.

\bibitem{Baltrusaitis2016}
T.~Baltru\v{s}aitis, P.~Robinson, and L.-P. Morency, ``Openface: an open source
  facial behavior analysis toolkit,'' in \emph{IEEE Winter Conference on
  Applications of Computer Vision}, 2016.

\bibitem{Lithari2010}
C.~Lithari, C.~A. Frantzidis, C.~Papadelis, A.~B. Vivas, M.~A. Klados,
  C.~Kourtidou-Papadeli, C.~Pappas, A.~A. Ioannides, and P.~D. Bamidis, ``Are
  females more responsive to emotional stimuli? a neurophysiological study
  across arousal and valence dimensions,'' \emph{Brain Topography}, vol.~23,
  no.~1, pp. 27--40, 2010.

\bibitem{wells2016identification}
L.~J. Wells, S.~M. Gillespie, and P.~Rotshtein, ``Identification of emotional
  facial expressions: Effects of expression, intensity, and sex on eye gaze,''
  \emph{PloS one}, vol.~11, no.~12, 2016.

\bibitem{koelstra2012fusion}
S.~Koelstra and I.~Patras, ``{Fusion of facial expressions and EEG for implicit
  affective tagging},'' \emph{Image and Vision Computing}, vol.~31, no.~2, pp.
  164--174, 2013.

\bibitem{Muhl14}
C.~Muhl, B.~Allison, A.~Nijholt, and G.~Chanel, ``A survey of affective brain
  computer interfaces: principles, state-of-the-art, and challenges,''
  \emph{Brain-Computer Interfaces}, vol.~1, no.~2, pp. 66--84, 2014.

\bibitem{halljudith}
J.~A. Hall and D.~Matsumoto, ``Gender differences in judgments of multiple
  emotions from facial expressions,'' \emph{Emotion}, vol.~4, no.~2, pp.
  201--206, 2004.

\bibitem{Mclure04}
E.~B. McClure, C.~S. Monk, E.~E. Nelson, E.~Zarahn, E.~Leibenluft, R.~M.
  Bilder, D.~S. Charney, M.~Ernst, and D.~S. Pine, ``A developmental
  examination of gender differences in brain engagement during evaluation of
  threat.'' \emph{Biological psychiatry}, vol.~55, pp. 1047--55, 2004 Jun 1
  2004.

\end{thebibliography}

\end{document}